\documentclass[aps,prl,floatfix,tightenlines,superscriptaddress]{revtex4}
\usepackage{amssymb, amsfonts, latexsym, graphicx, bbm, multirow, textcomp, amsmath}
\usepackage{color}
\usepackage[bookmarks = true, colorlinks = true, urlcolor=blue]{hyperref}

\begin{document}

\title{Direct generation of photon triplets using cascaded photon-pair sources}

\author{H.~H\"{u}bel}
\affiliation{Institute for Quantum Computing and Department of
Physics \& Astronomy, University of Waterloo, Waterloo, Canada,
N2L 3G1}
\author{D.R. Hamel}
\affiliation{Institute for Quantum Computing and Department of
Physics \& Astronomy, University of Waterloo, Waterloo, Canada,
N2L 3G1}
\author{A. Fedrizzi}
\affiliation{Department of Physics and Centre for Quantum Computer Technology and Department of Physics, University of Queensland, Brisbane, QLD 4072, Australia}
\author{S. Ramelow}
\affiliation{Institute for Quantum Optics and Quantum
Information (IQOQI), Austrian Academy of Sciences,
Boltzmanngasse 3, 1090 Vienna, Austria}
\affiliation{Faculty of Physics, University Vienna, Boltzmanngasse 5, 1090 Vienna, Austria}
\author{K.J. Resch}
\affiliation{Institute for Quantum Computing and Department of
Physics \& Astronomy, University of Waterloo, Waterloo, Canada,
N2L 3G1}
\author{T. Jennewein}
\affiliation{Institute for Quantum Computing and Department of
Physics \& Astronomy, University of Waterloo, Waterloo, Canada,
N2L 3G1}

\begin{abstract}
Non-classical states of light, such as entangled photon pairs and number states, are essential for fundamental tests of quantum mechanics and optical quantum technologies. The most widespread technique for creating these quantum resources is the spontaneous parametric down-conversion (SPDC) of laser light into photon pairs\cite{kly67}. Conservation of energy and momentum in this process, known as phase-matching, gives rise to strong correlations which are used to produce two-photon entanglement in various degrees of freedom\cite{ou88,shih88,rarity90,kwiat95,brendel99,mair01,barreiro05,ram09}. It has been a longstanding goal of the quantum optics community to realise a source that can produce analogous  correlations in photon \emph{triplets}, but of the many approaches considered, none have been technically feasible\cite{rarity98,persson04,greenberger90,keller98,gupta54,douady04,bencheikh07,guo05}.
In this paper we report the observation of photon triplets generated by cascaded down-conversion. Here each triplet originates from a \emph{single} pump photon, and therefore quantum correlations will extend over all three photons\cite{munro98} in a way not achievable with independently created photon pairs\cite{zukowski95}.  We expect our photon-triplet source to open up new avenues of quantum optics and become an important tool in quantum technologies. Our source will allow experimental interrogation of novel quantum correlations\cite{banaszek97}, the post-selection free generation of tripartite entanglement\cite{greenberger90, Zeilinger92} without post-selection and the generation of heralded entangled-photon pairs suitable for linear optical quantum computing\cite{Browne05}.  Two of the triplet photons have a wavelength matched for optimal transmission in optical fibres, ideally suited for three-party quantum communication\cite{hillery99}. Furthermore, our results open interesting regimes of non-linear optics, as we observe spontaneous down-conversion pumped by single photons, an interaction also highly relevant to optical quantum computing.
\end{abstract}

\maketitle

Given the potential for fundamental and applied quantum sciences,
several physical systems have been proposed for the direct generation of photon triplets. These include four-level atomic cascades  and higher-order optical nonlinearities\cite{rarity98}, tri-excitons in quantum dots\cite{persson04}, combinations of second-order nonlinearities\cite{keller98}, and high-energy electron-positron collisions\cite{gupta54}. Extremely low interaction strengths and collection efficiencies have rendered these proposals unfeasible. Recent experiments have observed and studied third-order\cite{douady04,bencheikh07} and cascaded second-order nonlinear\cite{guo05} parametric processes seeded by strong lasers. However, such seeding only increases stimulated emission which masks the production of three-partite quantum correlations and cannot lead to three-photon entanglement.

\begin{figure}
\includegraphics[width=10cm]{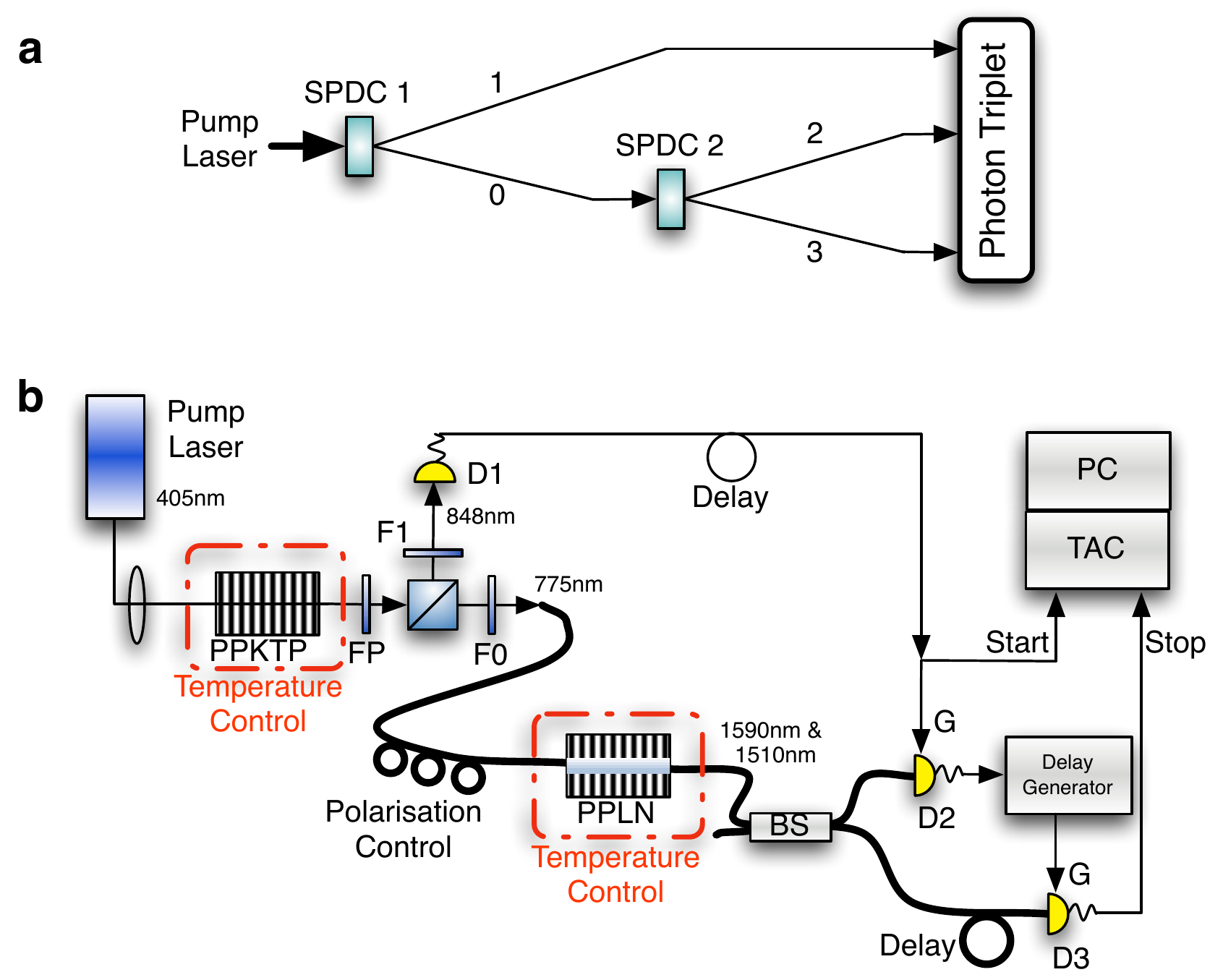}
\caption{ \textbf{Schematic of photon triplet generation and experimental setup.} \textbf{a}, A down-conversion source (SPDC1) produces a pair of photons in spatial modes 0 and 1, where the photon in mode 0 creates another photon pair in the second source (SPDC 2) in modes 2 and 3, generating a photon triplet. . \textbf{b}, The primary source, pumped by a 405~nm laser, produces photon pairs at 775~nm and 848~nm. The 848~nm photon is directly detected by a silicon avalanche photo-diode (D1), while the 775~nm photon serves as input to the secondary source, creating a photon pair at 1510~nm and 1590~nm, which is detected by two InGaAs avalanche photo-diodes (D2 and D3). A detection event at D3 represents a measured photon triplet. \label{setup}}

\end{figure}

Production of photon triplets by cascaded spontaneous parametric down-conversion (C-SPDC) was first proposed 20 years ago\cite{greenberger90}, yet never experimentally realised. The basic idea is shown in Fig.~\ref{setup}a. A primary down-conversion source is pumped by a laser to create a photon pair. One of the photons from this pair drives a secondary down-conversion process, generating a second pair and hence a photon triplet. Since the photon-triplet originates from a single pump photon, the created photons have strong temporal correlations\cite{Burnham70} and their energies and momenta sum to those of the original photon.

The C-SPDC process can be described using a simplified quantum optical model. The interaction Hamiltonian for the primary source can be written as, $H_1 = \lambda_1 \alpha (a_0^\dagger a_1^\dagger + h.c. )$, with the pump laser treated as a classical field with amplitude $\alpha$, and the photon creation operators of the two output modes $a_0^\dagger$ and $a_1^\dagger$. The coupling strength between the interacting fields is expressed by the parameter $\lambda_1 $, which includes the nonlinear response of the material and governs the expected conversion rate of pump photons. For the second down-conversion, the pump field is now a single photon and must be treated quantum mechanically in the interaction Hamiltonian, $H_2 = \lambda_2 (a_0  a_2^\dagger a_3^\dagger + h.c. )$, with output modes 2 and 3. The evolution operator of the system is $U = U_2 U_1 = \exp (-i H_2)  \exp (-i H_1) $, and can be approximated by expanding each term to first order. Applying $U$ to the initial vacuum state and ignoring the vacuum contribution for the final state results in 
\begin{eqnarray}
|\Phi\rangle & = & U |0_0,0_1,0_2,0_3\rangle  \label{eq:state} \\
& \approx&     - i \lambda_1 \alpha |1_0,1_1,0_2,0_3\rangle  - \lambda_1 \lambda_2 \alpha |0_0,1_1,1_2,1_3\rangle, \nonumber 
\end{eqnarray}
\noindent where the subscripts label the spatial modes. The first term describes the pair creation process in the first crystal, while the second represents the desired three-photon state, $ |0,1,1,1\rangle $, where the amplitude scales as the product of the two coupling strengths $\lambda_{1}$ and $\lambda_{2}$ of both down-converters. Note that Eq. \ref{eq:state} predicts that the rate of triplet production from C-SPDC should be linear in the intensity of the pump laser.

The conversion efficiencies in SPDC are typically very low. In optical nonlinear materials such as $\beta$-Barium Borate, for example, they reach about $10^{-11}$ per pump photon\cite{kurtsiefer01a}. Major advances in nonlinear optics, such as quasi-phasematching of optical materials, have recently made it possible to access the inherent higher nonlinearities of materials such as periodically-poled Lithium Niobate (PPLN) and periodically-poled Potassium Titanyl Phosphate (PPKTP). The down-conversion efficiencies demonstrated in these materials can reach up to $10^{-9}$ in bulk\cite{fedrizzi07}. The introduction of optical waveguides in photon-pair sources\cite{Tittel01a} has further increased conversion efficiencies to $10^{-6}$, making the observation of C-SPDC possible. 

Figure~\ref{setup}b depicts the experimental setup (see Methods for more details). The primary source generated photon pairs in a PPKTP crystal, quasi-phasematched for collinear SPDC of 405~nm $\rightarrow$ 775~nm + 848~nm. The 775~nm photons were used to pump the secondary source, consisting of a PPLN waveguide, quasi-phasematched for 775~nm $\rightarrow$ 1510~nm + 1590~nm. The photon triplets were measured using a chained series of three photon counters (D1, D2 and D3) based on avalanche photo-diodes (APD). The detection of a 848~nm photon at D1, which occurred at a rate of about 1~MHz, opened a 20~ns gate at D2, which in turn gated D3 for 1.5~ns. The actual gate rate of D2 was reduced to 870~kHz, due to saturation. Since D3 was only armed if both D1 and D2 had fired, an event at D3 constituted the detection of a photon triplet. The temporal signature of these triple coincidences was recorded as histograms with a fast time  acquisition card, where the detection signal of D1 served as the start trigger, and the detection signal on D3 as the stop. Data were recorded for a total of 20 hours, and analysed as a histogram of the time-interval between D3 and D1 detections, $\Delta\tau_{D3-D1}$. 

\begin{figure}
\includegraphics[width=10cm]{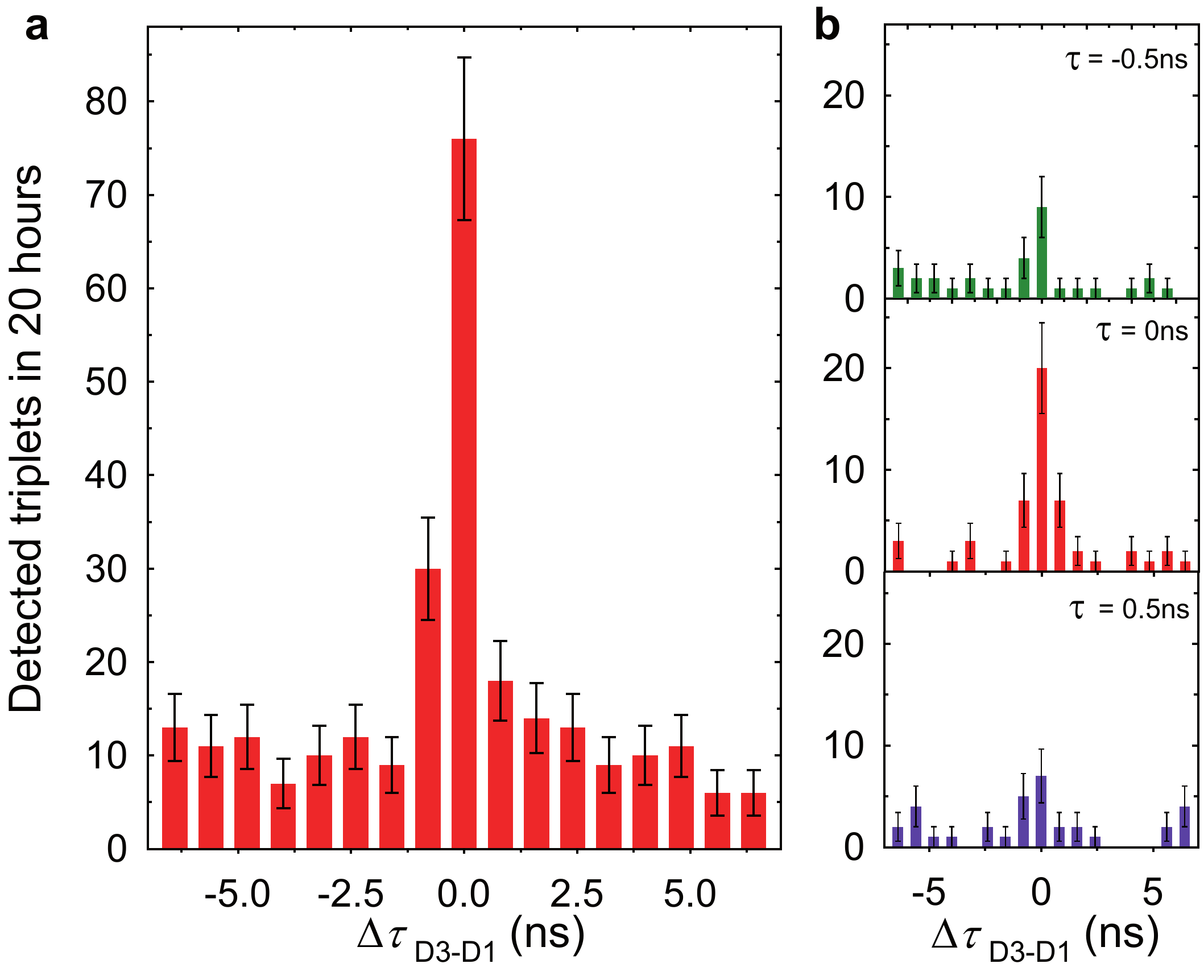}
\caption{\textbf{Triple-coincidence histograms.} \textbf{a}, Measured triple coincidences obtained in 20 hours. Each bin corresponds to a 0.8~ns time interval between events at D3 and D1 ($\Delta\tau_{D3-D1}$). The sharp peak indicates a strong temporal correlation between all three detection events, as expected of the C-SPDC process. \textbf{b}, Triple-coincidence histograms with varying delays of $\tau=0$~and~$\pm \,0.5$~ns between D2 and D3, resulting in a decrease of the coincidence peak. Note that the absolute rate reduction for $\tau=0$ results from a different setting on the InGaAs detectors for this measurement series. Error bars represent one standard deviation. \label{hist}}

\end{figure}

A typical data set, shown as a histogram in Fig.~\ref{hist}a, displays a peak 8 standard deviations above the background noise. This is a clear signature of C-SPDC photon triplets. The 1.2~ns temporal width of the observed photon-triplet peak is dominated by detector jitter. Integration over the three central time bins yields a raw triplet rate of 124~$\pm$~11 events in 20 hours. The observed background in the histogram is caused predominantly by triple events between a genuine detection in D1 and dark counts in D2 and D3 (see Methods) and was estimated from the displayed data to be 10.2~$\pm$~0.9 per bin in 20 hours. The detected rate of triplets, exclusively produced by the C-SPDC process, was 4.7~$\pm$~0.6 counts per hour. We modelled the process under the assumption that the down-conversion efficiency per photon in the secondary source was independent of the pump intensity (see Supplementary Information). Using the conversion efficiencies obtained from independent characterisations of both sources at mW pump power, and optical parameters from other relevant components of our setup, our model predicts a triplet rate of 5.6~$\pm$~1.1 counts per hour, which is in very good agreement with the measured value.

It is expected that C-SPDC photon triplets should exhibit strict time correlations\cite{Burnham70}. We investigated this property by introducing three different delays between D2 and D3 (-0.5, 0 and 0.5 ns) and measuring the histograms.  The data in Fig.~\ref{hist}b  shows a significant reduction of the peak in the histograms with additional delays, verifying the strong temporal correlations of the created triplets.

\begin{figure}
\includegraphics[width=10cm]{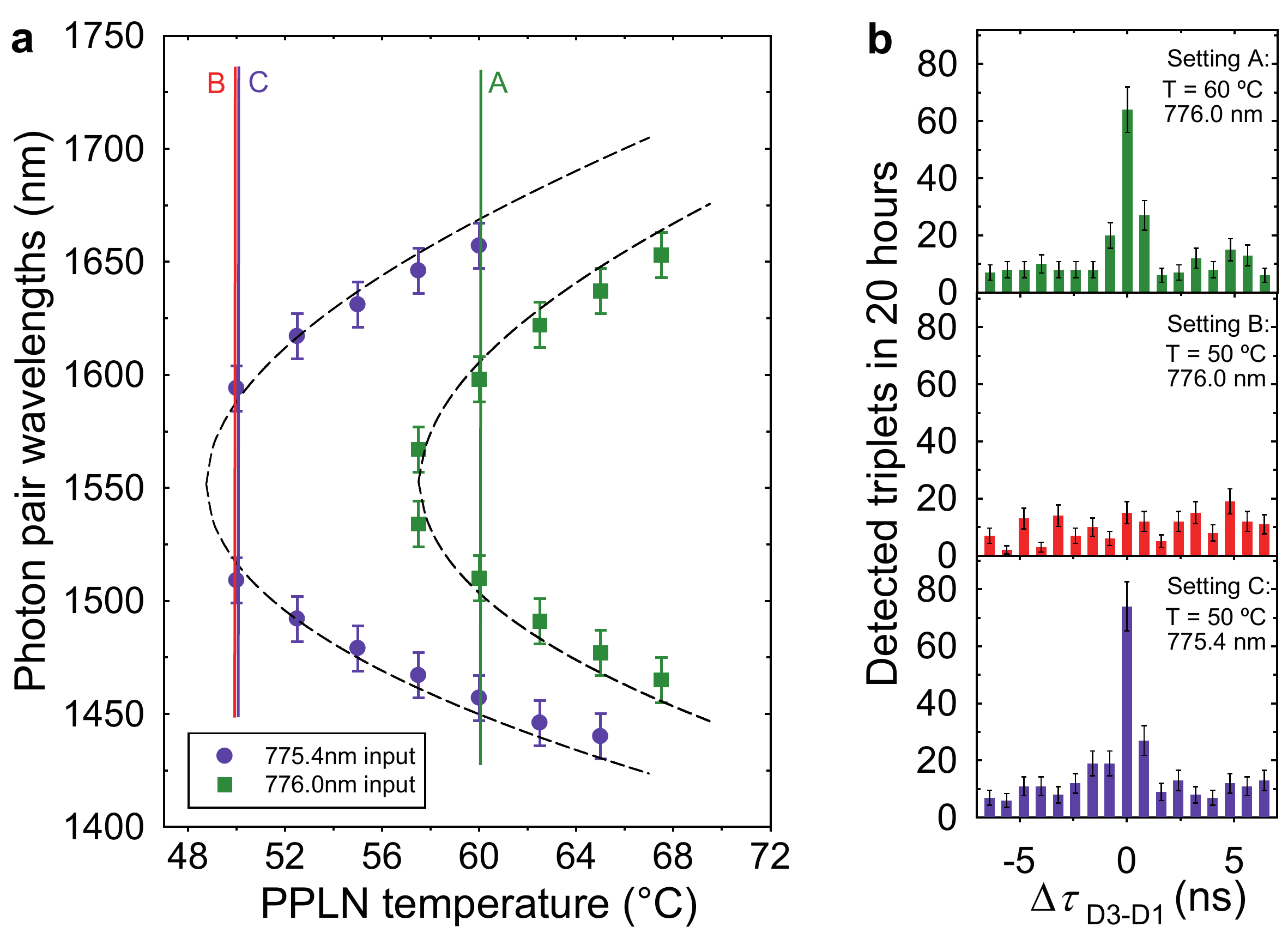}
\caption{\textbf{Phase-matching and triple-coincidence dependence on crystal temperatures.} \textbf{a}, Central wavelengths of the pair of photons produced by the secondary source as a function of the PPLN temperature for input wavelengths of 775.4~nm (circles) and 776.0~nm (squares). The dashed line shows the theoretic phase-matching curve with the  poling period as the only  fit parameter. Triple coincidences were measured for different settings of the PPLN temperature and the input pump photon wavelength. The PPLN temperature was 60 °C for setting A and 50 °C for settings B and C; the input photon wavelength was 776.0 nm for settings A and B and 775.4 nm for setting C.  \textbf{b}, Measured triple coincidence histograms over 20 hours for each measurement setting. For A and C, the PPLN temperatures lie on the respective phase-matching curves and a triple coincidence peak is observed. For B, the temperature is outside the 776.0~nm phase-matching curve and no peak is present. Wavelength changes in the input photons, needed for the measurements shown in Fig.~\ref{tuning}b, were achieved by altering the temperature of the PPKTP crystal ($43.6^\circ$C for setting A and B, and $40.8^\circ$C for C). Error bars represent one standard deviation. \label{tuning}}

\end{figure}

It is conceivable that other physical processes, such as the APD breakdown flash from D1\cite{kurtsiefer01b}, electronic cross-talk, or double-pair emission from the primary source, might give rise to correlated triple detection events with similar features to the ones we have observed. We can  rule out these alternatives by testing the expected dependence of the C-SPDC signal on temperature and input wavelength of the secondary down-conversion. As shown in Fig.~\ref{tuning}a, for a given input wavelength into the PPLN crystal, phase-matching imposes a minimum temperature below which down-conversion cannot occur. The triple coincidence peak, in Fig.~\ref{tuning}b, indeed disappears when the PPLN temperature is lowered from 60$^\circ$C (setting A) to 50$^\circ$C (setting B), while keeping the input wavelength fixed at 776.0~nm. The triple photon signal is then recovered at this temperature, by lowering the input wavelength to 775.4~nm (setting C). These measurements, together with the strong agreement between the observed and predicted triplet rate, provide conclusive proof that we have indeed observed spontaneously produced photon triplets. 

In the near future we expect to increase the photon-triplet rate by at least one order of magnitude using an improved time acquisition system, a dichroic beamsplitter for separating the photons created in the secondary source and matching the down-conversion bandwidth of the initial pair to the PPLN crystal. The direct generation of the triplet guarantees strong energy-time correlations, allowing the creation of entangled, or hyper-entangled\cite{Kwiat97} triplets and  realisations of tri-partite states like GHZ\cite{bouwmeester99} and W\cite{kiesel03} without elaborate and probabilistic post-selection schemes. For example, time-bin entangled\cite{brendel99} GHZ states could be produced by pumping our triplet source with a pulsed pump laser in a coherent superposition of two time slots. The entanglement could then be detected using three standard unbalanced interferometers. For a further example, W-states could be made by using an entangling source as the primary down-converter producing a Bell-state $\frac{1}{\sqrt{2}}(|V_0V_1\rangle + |H_0H_1\rangle)$, where $|V\rangle$ and $|H\rangle$ denote the photon polarization states in their respective modes. The secondary source would consist of two down-converters where $|V_0\rangle$ is converted to $|H_2H_3\rangle$, and $|H_0\rangle$ is converted to $\frac{1}{\sqrt{2}}(|H_2V_3\rangle + |V_2H_3\rangle)$, into the same pair of modes. The relative amplitudes could then be balanced by tuning the conversion efficiencies. Polarization entangled GHZ states could be made by modifying the W-state scheme such that the secondary source converts $|V_0\rangle$ to $|V_2V_3\rangle$ and $|H_0\rangle$ to $|H_2H_3\rangle$. An interesting application of such a GHZ source could be to herald the presence of an entangled photon pair in mode 1 and 2 by detecting the secondary down-converted photon in mode 3. This has proven very difficult to achieve otherwise. Our results also confirm that the SPDC conversion efficiency is independent of pump power down to the single photon level (see Supplementary Information), allowing new tests of nonlinear optics in the quantum regime.

\subsection{Acknowledgments}
The authors would like to thank H. Majedi and G. Weihs for providing equipment and infrastructure for implementing the experiment. Gratefully acknowledged is the financial support by the Canadian Institute for Advanced Research, the Ontario Centres of Excellence, the Ontario Ministry of Research and Innovation, the Natural Sciences and Engineering Council of Canada and the Canadian Foundation for Innovation. S.R. acknowledges support from the FWF (CoQus).
 H.H. and D.R.H. performed the experiment and analysed the data; A.F. and S.R. participated in the conception of the experiment;  K.J.R. and T.J. contributed to the conception and realisation of the experiment; and all authors co-wrote the paper.
The authors declare that they have no competing financial interests.  Correspondence and requests for materials should be addressed to H.H. or T.J~(email: hhuebel@iqc.ca, tjennewe@iqc.ca).

\section{Methods}

\textbf{Experimental setup.} The primary source, shown in Fig.~\ref{setup}b, consisted of a 25~mm long, temperature-stabilised  PPKTP crystal and was pumped with 2.4~mW from a 405~nm continuous-wave diode laser. The type-II SPDC in the PPKTP generated orthogonally-polarised photons at 775~nm and 848~nm, which were separated by a polarising beamsplitter and coupled into single mode fibers. A longpass filter (FP) was used to block the strong 405~nm pump, passband filters (12~nm bandwidth) with central wavelengths of 780~nm (F0) and 840~nm (F1) respectively, were placed before the fiber couplers to further reduce background.  The 775~nm photon, after passing an in-fiber polarisation controller, served as input to the secondary source, a 30~mm temperature-stabilised PPLN waveguide crystal with fiber pigtails attached to both ends for type-I SPDC. The photon pair at 1510~nm and 1590~nm was separated using a  50:50 fiber beamsplitter (BS). The secondary source was operated without filters, as the input power during a C-SPDC measurements was low enough ($\sim 1$ million input photons per second) not to cause additional detection events in the InGaAs detectors. The gate (G) and photon arrivals on these detectors were synchronised by an internal delay generator on D2, and an external delay generator between D2 and D3. Detection efficiencies on the InGaAs detectors D2 and D3 were set to 20\% and 10\%, respectively. Trigger events from D1 and detection events from D3 were recorded via a time acquisition card (TAC) with a timing resolution of 103~ps and analysed on a computer (PC). 

\textbf{Dark count rate.} The total background  during the 20 hour runs, seen in Fig.~\ref{hist}a, was measured to be 268~$\pm$~16 events over the whole 20~ns gate. This number is in very good agreement with the expected noise count of 254~$\pm$~5 triple events as calculated from the individual dark count probabilities per gate of D2 and D3 ($1.8 \times 10^{-3}$ and $4.5 \times 10^{-6}$), trigger rate and efficiency of the time acquisition card.

\section{Supplementary Information for ``Direct generation of photon triplets using cascaded photon-pair sources''}
\subsection{Calculation of the expected triplet detection rate}

The predicted triplet rate, $R_{triple}$, quoted in our manuscript was calculated in the following way. Considering the efficiencies of all the components of the C-SPDC setup and the down-conversion probability, the triple coincidence probability ($P_{triple}$) per trigger at D1, is given by
\begin{equation}
\label{ptriple1}
P_{triple}=\eta_{775} \: \eta_{in} \: P_{SPDC} \: \eta_{out}^2 \: 2\,\eta_{BS}^2 \: \eta_{D2} \: \eta_{D3}
\end{equation}
where $\eta_{775}$ is the fiber coupling probability of the
primary SPDC, $\eta_{in}$ and $\eta_{out}$ are the coupling
efficiencies into and out of the waveguide respectively, $P_{SPDC}$ is the intrinsic
downconversion probability of the PPLN waveguide, $\eta_{BS}$ is the
transmission of the fiber beamsplitter for either arms, and
$\eta_{D2}$ and $\eta_{D3}$ are the detection
efficiencies of the InGaAs-APDs D2 and D3, respectively.

The probability, $\eta_{775}$, to find a 775~nm photon in the
output fiber of the primary source upon detection of its partner
photon at 848~nm was estimated from a coincidence measurement
between the 775 and 848~nm photons, where a coincidence to
singles ratio of 0.24 was measured. Assuming a detection
efficiency of $\eta_{D1}=$~0.45~$\pm$~0.05 for the Si-APD (D1), the fiber coupling probability, $\eta_{775}$, is 0.53~$\pm$~0.06.

The probability, $P_{coinc}$, of a detected coincidence event between the pair photons of the secondary source, given a single 775~nm photon in the input
fiber of the waveguide, can be written as:
\begin{equation}
\label{coinc}
P_{coinc}= \eta_{in}  \: P_{SPDC} \: \eta_{out}^2 \: \eta_{LP}^2 \: 2\,\eta_{BS}^2 \: \eta_{D2} \: \eta_{D3} \: \eta_{duty}
\end{equation}

\noindent where $\eta_{LP}$ is the transmission of a longpass filter needed
to block the strong pump light and
$\eta_{duty}$ is the duty cycle of detector D2,
which was operated in a quasi free running mode with 100~ns gate
width and 100~kHz gate repetition rate.
$P_{coinc}$ was measured, using a cw-laser with 245~nW power, and 24~$\pm$~2
coincidences per second were observed. Converting the pump power into number of photons per second
numbers ($9.56 \times 10^{11}$~s$^{-1}$) we arrive at a coincidence
probability, $P_{coinc}$, of $(2.5\pm0.2) \times 10^{-11}$.

Combining Eq. \ref{ptriple1} and
\ref{coinc} leads to:
\begin{equation}
\label{ptriple2}
P_{triple}=\eta_{775} \:  \frac{P_{coinc}}{\eta_{LP}^2 \: \eta_{duty}}
\end{equation}
To arrive at the observed triplet rate, the trigger rate ($R_{trigger}$), and experimental values for the efficiency of the time acquisition card ($\eta_{TAC}$) and the SPDC bandwidth dependence ($\eta_{cw}$) have to be included, finally yielding:
\begin{equation}
\label{ptriple3}
R_{triple}=R_{trigger} \: \eta_{775} \:  \frac{P_{coinc}}{\eta_{LP}^2 \: \eta_{duty}} \: \eta_{TAC} \: \eta_{cw}
\end{equation}
Due to the internal deadtime, the time acquisition card only counted every second event and hence reduced the number of observed events by a factor of 2. Measurements of the PPLN waveguide have also shown that
the acceptance of the down-conversion process with respect to
the input wavelength is very small. The overall efficiency is hence reduced when broadband input photons (0.4~nm) are used as was the case in the C-SPDC measurements. The efficiency with such a broad pump was calculated to be $73\%$ for optimal
matching of the single photon wavelength and dropping to 55\%
when the input wavelength is 0.2~nm off. For the calculation of $R_{triple}$, a value of 0.67~$\pm$~0.05 for $\eta_{cw}$ is assumed here.
Substituting the experimental values, as summarised in Table~\ref{tab1}, in Eq.~\ref{ptriple3}, a triplet rate of 5.6~$\pm$~1.1 counts per hour is found.

\subsection{Measurement of the down-conversion efficiency in the PPLN waveguide}

By using the estimates for the various losses and efficiencies, the down-conversion probability per pump photon inside the waveguide of the secondary SPDC source, $P_{SPDC}$, was calculated from a coincidence measurement (Eq.~\ref{coinc}) using 245~nW laser power and also from the measured triplets rate (Eq.~\ref{ptriple1}) where $10^6$ single photons per second were used. The laser pumping yields a $P_{SPDC}$ of $(9.9\pm2.9)\times 10^{-6}$, compared to $(8.2\pm2.2)\times 10^{-6}$ for the single photon input. The down-conversion efficiency was also measured at a higher laser power of 1.1~mW. For this input, an SPDC power of 0.9~$\pm$~0.1~nW could be detected at the output fiber of the waveguide using a standard power meter. Including the losses of the fiber couplings and the longpass filter, this measurement yields an $P_{SPDC}$ of $(6.6\pm 0.7) \times 10^{-6}$ for the waveguide.

The agreement, within measurement errors, of the down-conversion efficiency obtained at input powers of 1.1~mW and 245~nW and for the single photon input is very good. This leads us to the conclusion that the SPDC probability is indeed constant over the observed power range, from 260~fW ($10^6$ pump photons per second) to 1~mW ($10^{15}$ pump photons).

\begin{table}
\centering
\begin{tabular}{c c }
\hline
$R_{trigger}$ & ($8.70\pm0.05)\times 10^{5} \mathrm{Hz}$\\
$\eta_{D1}$ & $0.45\pm0.05$ \\
$\eta_{775}$ & $0.53\pm0.06$ \\
$P_{coinc}$ & $(2.5\pm0.2)\times 10^{-11}$ \\
$\eta_{LP}$ & $0.50\pm0.03$ \\
$\eta_{duty}$ & $0.01$ \\
$\eta_{TAC}$ & $0.5$ \\
$\eta_{cw}$ & $0.67\pm0.05$\\
$\eta_{in}$ & $0.50\pm0.05$\\
$\eta_{out}$ & $0.50\pm0.05$\\
$\eta_{BS}$ & $0.45\pm0.05$\\
$\eta_{D2}$ & $0.20\pm0.02$\\
$\eta_{D3}$ & $0.10\pm0.01$\\
\hline
\end{tabular}
\caption{Experimental parameters used for the calculation of the triple coincidence rate ($R_{triple}$) and PPLN down-conversion efficiency ($\eta_{SPDC}$). Errors correspond to $1\sigma$.}
\label{tab1}
\end{table}

\end{document}